\documentclass[prx,preprint,showpacs,preprintnumbers,amsmath,amssymb,floatfix]{revtex4}

\usepackage[dvipdfmx]{graphicx}
\usepackage{bm}

\newcommand{\be}{\begin{equation}}
\newcommand{\ee}{\end{equation}}
\newcommand{\eq}[1]{Eq.~(\ref{#1})}
\newcommand{\fig}[1]{Fig.~\ref{#1}}
\def\bea{\begin{eqnarray}}
\def\eea{\end{eqnarray}}

\def\bra{\langle}
\def\ket{\rangle}
\def\vq{{\bf q}}

\def\vk{{\bf k}}
\def\vQ{{\bf Q}}

\begin{document}

\title{Spin-fluctuation glue disfavors high-critical temperature of superconductivity?} 

\author{Hiroyuki Yamase} 
\affiliation{
International Center of Materials Nanoarchitectonics, National Institute for Materials Science, Tsukuba 305-0047, Japan\\
}
\date{December 9, 2022}

\begin{abstract}
Antiferromagnetic fluctuations are believed to be a promising glue to drive high-temperature superconductivity  especially in cuprates. Here, we perform a close inspection of the superconducting mechanism from spin fluctuations in the Eliashberg framework by employing a typical one-band model on a square lattice.  While spin fluctuations can eventually drive superconductivity as is well established, we find that the superconducting tendency is suppressed substantially by a seemingly negligible contribution from a small momentum transfer far away from $(\pi,\pi)$. This suppression comes from phase frustration of the pairing gap and is expected to be a general feature due to the repulsive pairing  interaction of spin fluctuations. Furthermore, we find that the momentum dependence of the pairing gap largely deviates from  the functional form of $\cos k_{x} - \cos k_{y}$, although this form is well established in cuprate superconductors. We argue that an instantaneous magnetic interaction plays the important role to understand high-critical temperature of superconductivity as well as the momentum dependence of the pairing gap. 
\end{abstract}


\maketitle
\section{introduction}
The mechanism of superconductivity has been a central issue in condensed matter physics since the discovery of superconductivity in mercury \cite{onnes1911}. The crucial role of electron-phonon coupling was successfully formulated microscopically in the BCS theory \cite{bardeen57,schrieffer}. The phonon dynamics is responsible for the retardation effect in the Cooper pairing and this effect was formulated in the Eliashberg theory \cite{eliashberg60,schrieffer,marsiglio20}. Phonon-mediated superconductivity was widely discussed as conventional  superconductors typically with low critical temperature ($T_{c}$) for a long time. A large number of analyses in terms of the Allen-Dynes formula \cite{mcmillan68,allen75} suggested that the highest $T_{c}$ could be around 30 K, which seemed valid among many superconductors. This insight was challenged by the discovery of MgB$_{2}$ with $T_c = 39$ K in 2001 \cite{nagamatsu01,an01}. More recently, metal hydrides were shown to exhibit $T_{c}$ with 200--260 K under high pressure 150--200 GPa \cite{drozdov15,drozdov19,somayazulu19}. The observation of an isotope effect \cite{drozdov15,drozdov19} suggested that the hydride superconductors are driven by the conventional electron-phonon coupling mechanism and regarded as an experimental realization of a theoretical prediction by Ashcroft \cite{ashcroft68,ashcroft04}. 

Superconductivity is also stabilized near a magnetic phase. Naturally spin fluctuations were argued as a mechanism of superconductivity in cuprates, iron-based pnictides and chalcogenides, and various heavy electron materials \cite{scalapino12}, as well as organic conductors \cite{mckenzie97}. Theoretically, the Eliashberg framework was frequently taken over to study superconductivity in those systems by replacing a phonon spectrum with a spin excitation spectrum. One may also replace it with other bosonic fluctuations. For example, in iron-based superconductors, various bands with different orbital characters cross the Fermi energy. In this sense, orbital fluctuations may also be responsible for superconductivity \cite{stanescu08,kontani11}. A close look at the phase diagram in iron-based superconductors reveals that the so-called electronic nematic phase surrounds the magnetic phase in many cases \cite{stewart11} and in this sense the superconducting phase is realized closer to the nematic phase than the magnetic phase. Nematic fluctuations were also discussed as a possible superconducting mechanism \cite{yamase13b,agatsuma16}.   

When the superconductivity is mediated by bosonic fluctuations such as phonons, spin fluctuations, and others, the retardation effect plays the crucial role so that two electrons can make a pair with a time delay. Mathematically, the retardation effect is described by the energy dependence of bosonic fluctuations. On the other hand, the bosonic fluctuations can contain a term independent of energy. Such a term describes a pairing interaction between two electrons that propagates instantaneously---no time delay. Since the physical property of the instantaneous interaction is sharply different from the retarded interaction, we distinguish between them in the present work. A pairing interaction due to the retardation effect is frequently referred to also as a {\it glue} of superconductivity, but here we use the former term to emphasize the contrast to the instantaneous interaction.  Both retarded and instantaneous interactions can be considered in many studies, but usually the retardation effect, i.e., a frequency-dependent part is paid attention to much more. Following this convention, in this paper, we mean the Cooper pairing due to the retarded magnetic interaction by the spin-fluctuation mechanism of superconductivity. An important, but much less addressed,  question is why spin fluctuations can achieve high $T_{c}$ in cuprates, given that spin fluctuations are also believed to be a promising mechanism of superconductivity in heavy electron materials \cite{scalapino12}, where $T_{c}$ in absolute value is usually much lower than $T_{c}$ in cuprates. 

On the other hand, Anderson discussed the importance of an instantaneous interaction as the superconducting mechanism of cuprates---a glue is {\it not} necessary to high-$T_{c}$ cuprate superconductors \cite{anderson07}.  The instantaneous interaction is given by the superexchange interaction $J=4 t^{2}/U$, which is a direct consequence of the large on-site Coulomb repulsion $U$; $t$ is the nearest-neighbor hopping integral.  This insight was supported by coherent charge fluctuation spectroscopy \cite{mansart13} and a recent scanning tunneling microscopy \cite{omahony22} in cuprates. In addition, Ref.~\onlinecite{dong22} reported that only about half of the pairing can be attributed to spin fluctuations in the Hubbard model with $U/t=6$. However, other studies of the $t$-$J$ \cite{prelovsek05,maier08} and Hubbard \cite{maier08,kyung09,dong22a} models suggested a dominant role of spin fluctuations, namely the retardation effect, in the superconducting mechanism. 

In this paper, aiming to gain a general view to achieve possibly high $T_{c}$, including the interplay of retarded and instantaneous interactions, we perform a close inspection of the superconducting mechanism in the Eliashberg theory. The key aspect is phase frustration of the pairing gap induced by spin fluctuations. We find that while the superconductivity is eventually driven by spin fluctuations characterized by momentum transfer $(\pi,\pi)$, the transition temperature is suppressed nearly by half owing to the {\it tail} of the spin fluctuation spectrum, which generates phase frustration of the pairing gap.  We call this the {\it self-restraint effect} of superconductivity. We also find that while the pairing gap is characterized by $d$-wave symmetry, it largely deviates from the simple form such as $\cos k_{x} - \cos k_{y}$. The self-restraint effect can be circumvented by an instantaneous magnetic interaction, which boosts the critical temperature of superconductivity substantially with the simple $d$-wave pairing gap.

\section{Model and Formalism}
We study the superconducting instability from spin fluctuations on a square lattice system. Our Hamiltonian consists of two terms, 
\be
H = \sum_{\vk} \xi_{\vk} c^\dagger_{\vk \sigma} c_{\vk \sigma} + 
\frac{1}{8N}\sum_{\vk \vk' \vq} \sum_{\sigma_{i}} V(\vk, \vk', \vq) \boldsymbol{\sigma}_{\sigma_{1} \sigma_{2}} \cdot 
\boldsymbol{\sigma}_{\sigma_{3} \sigma_{4}} c^\dagger_{\vk \sigma_{1}} c_{\vk+\vq \sigma_{2}} 
c^\dagger_{\vk' + \vq \sigma_{3}} c_{\vk' \sigma_{4}} \,.  
\label{model}  
\ee
The first term describes kinetic energy of electrons with a dispersion 
\be
\xi_{\vk} = -2 t (\cos k_{x} + \cos k_{y}) - 4 t' \cos k_{x} \cos k_{y} -\mu \,,
\label{xik}
\ee
where $t$ and $t'$ are the first and second nearest-neighbor hopping; $\mu$ is the chemical potential.  $c^{\dagger}_{\vk \sigma}$ and $c_{\vk \sigma}$ are creation and annihilation operators of electrons with momentum $\vk$ and spin $\sigma$, respectively. The second term is a general SU(2) symmetric two-particle interaction---it describes the effective spin-spin interaction of itinerant electrons and is justified close to a spin-density-wave (SDW) instability; $\boldsymbol{\sigma}$ are Pauli matrices and $N$ is the total number of lattice sites. This interaction was employed to study superconductivity from spin fluctuations in various systems \cite{nakajima73,miyake86,moriya90}. Microscopically, it is obtained as a low-energy effective magnetic interaction generated by, for example, the repulsive Hubbard interaction in a functional renormalization group study \cite{husemann09,eberlein14}. Note that if we take the limit of $\vq  \rightarrow {\bf 0}$,  the interaction is reduced to the well-known SU(2)-symmetric Landau interaction function in the spin-antisymmetric channel. The  precise form of $V(\vk, \vk', \vq)$ depends on microscopic details. Since we consider the conventional SDW state described by the order parameter such as $\bra  {\bf S} (\vq) \ket = \frac{1}{2} \sum_{\vk, \alpha, \beta} \bra c^{\dagger}_{\vk \alpha} \boldsymbol{\sigma}_{\alpha \beta}  c_{\vk+\vq \beta} \ket$, we may approximate $V(\vk, \vk', \vq) \approx V(\vq)$ without losing a practical importance. As a functional form of $V(\vq)$, we consider $V(\vq) = 2 V ( \cos q_{x} + \cos q_{y})$, which favors a SDW state characterized by the momentum $\vq= (\pi, \pi)$ and may cover many interesting cases. In real space, this interaction means that the nearest-neighbor spin interaction is dominant. See Appendix~D for other choices of $V(\vq)$. 

We study the superconducting instability in the Eliashberg theory \cite{eliashberg60,schrieffer,marsiglio20}. The Eliashberg gap equations are described by two coupled equations for the pairing gap $\Delta (\vk, {\rm i} k_{n})$ and the renormalization function $Z (\vk, {\rm i} k_{n})$; ${\rm i} k_{n}= {\rm i} (2n+1) \pi T$ is Matsubara frequency at temperature $T$. It is well known that a numerical analysis of the Eliashberg gap equations is demanding and thus $Z$ is set to unity in many cases. Yet, calculations are often limited to a temperature region much higher than the actual $T_{c}$ because the number of Matsubara frequency increases with decreasing temperature. To overcome these difficulties, we recall that the superconductivity is a phenomenon close to the Fermi surface and project momentum $\vk$ on the Fermi surface. Technically this is a standard approach in the conventional electron-phonon mechanism \cite{schrieffer,marsiglio20} as well as the spin-fluctuation mechanism \cite{radtke92,millis92}, but we keep a fine resolution of momentum along the Fermi surface to handle an anisotropy of magnetic interactions precisely and a resulting anisotropic gap formation. We achieve calculations down to very low temperatures, including the effect of the renormalization function  $Z$. 

The resulting linearized Eliashberg equations are obtained as  \cite{eliashberg60,schrieffer,marsiglio20} 
\bea
&& \Delta(\vk_{F}, {\rm i} k_{n}) Z(\vk_{F}, {\rm i} k_{n}) = - \pi T \sum_{\vk_{F}', n'} N_{\vk_{F}'} \frac{\Gamma_{\vk_{F} \vk_{F}'} ({\rm i}k_{n}, {\rm i}k_{n}')}{| k_{n}' |} \Delta(\vk_{F}', {\rm i}k_{n}')  \label{eliashberg1} \,, \\
&& Z(\vk_{F}, {\rm i} k_{n}) =1 - \pi T \sum_{\vk_{F}', n'} N_{\vk_{F}'} \frac{k_{n}'}{k_{n}} \frac{\Gamma^{Z}_{\vk_{F} \vk_{F}'} ({\rm i}k_{n}, {\rm i}k_{n}')}{| k_{n}' |} \label{eliashberg2} \,.
\eea
Here in contrast to the conventional treatment \cite{eliashberg60,schrieffer,marsiglio20}, we keep the momentum dependence---the Fermi surface is divided into many patches and each patch is specified by discrete momentum $\vk_{F}$ [see also \fig{gap}(b)]. $N_{\vk_{F}}$ is a momentum-resolved density of states on the Fermi surface patch specified by $\vk_{F}$. $\Gamma_{\vk_{F} \vk_{F}'} ({\rm i}k_{n}, {\rm i}k_{n}')$ is the pairing interaction and is computed  from the magnetic interaction in \eq{model} as 
\be
\Gamma_{\vk_{F} \vk_{F}'} ({\rm i}k_{n}, {\rm i}k_{n}') = - \frac{1}{4} \bra \hat{V}(\vk - \vk', {\rm i} k_{n} - {\rm i} k_{n}' ) 
+ 2 \hat{V}(\vk + \vk', {\rm i} k_{n} + {\rm i} k_{n}' ) \ket_{\vk_{F} \vk_{F}'} \,, 
\label{Gamma}
\ee
where the first term on the right-hand side comes from the longitudinal spin fluctuations---$\sigma^{z}$ component in \eq{model}--- and the second one from the transverse ones;  $\bra \cdots \ket_{\vk_{F} \vk_{F}'}$ indicates the average over the Fermi surface patches $\vk_{F}$ and $\vk_{F}'$. $\hat{V}(\vq, {\rm i} q_{m})$ is the spin fluctuation propagator and is computed in the random-phase approximation 
\be
\hat{V}(\vq, {\rm i}q_{m}) = V^{*}(\vq) - \frac{V(\vq) \chi_{0}(\vq, {\rm i} q_{m})V(\vq)}{1+V(\vq)\chi_{0}(\vq, {\rm i}q_{m})} \,,
\label{Vhat}
\ee
where $\chi_{0}(\vq, {\rm i}q_{m})$ describes a simple bubble diagram and is given by  the Lindhard function. The first term in \eq{Vhat}  does not depend on frequency and accounts for the instantaneous interaction. Here we have $V^{*}(\vq)=V(\vq)$. However, from a physical point of view, it may be renormalized by the Coulomb repulsion, which is also instantaneous, as we shall discuss later [see \eq{V*}].  The second term describes the retardation effect on the pairing. It is usually this term which is discussed in the spin-fluctuation mechanism of superconductivity \cite{scalapino12}. Similarly, at the same order of approximation,  $\Gamma^{Z}_{\vk_{F} \vk_{F}'} ({\rm i}k_{n}, {\rm i}k_{n}')$ is computed as  
\be
\Gamma^{Z}_{\vk_{F} \vk_{F}'} ({\rm i}k_{n}, {\rm i}k_{n}') = \frac{1}{4} \bra 3 \hat{V}(\vk - \vk', {\rm i} k_{n} - {\rm i} k_{n}' ) 
-  2 V(\vk - \vk') \ket_{\vk_{F} \vk_{F}'} \,. 
\ee
After evaluating \eq{eliashberg2}, we end up solving the eigenvalue equation \eq{eliashberg1} numerically. When the eigenvalue $\lambda$ exceeds unity, superconducting instability occurs and the eigenvector describes the gap structure.

\section{Results}
We shall present results for $t'/t=-0.25$ and $V/t=0.95$ as representative ones, except for \fig{phase} where the $V$ dependence of $T_{c}$ is studied. The electron density is fixed to 0.85. Since it is vital to distinguish between the retarded and instantaneous interactions---second and first terms on the right hand side in \eq{Vhat}, we first study the former and then the later. We measure all quantities with the dimension of energy in units of $t$.   

\subsection{Self-restraint effect of superconductivity}
\begin{figure}[ht]
\centering
\includegraphics[width=7cm]{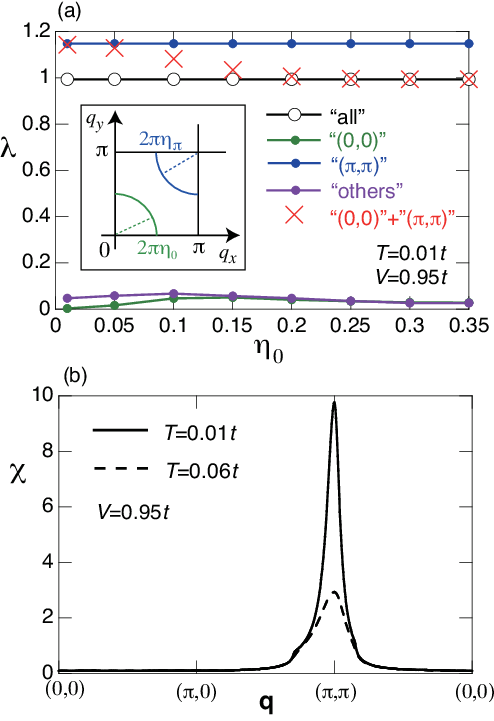}
\caption{(a) $\eta_{0}$ dependence of eigenvalue $\lambda$ for different scattering processes at $T=0.01(\approx T_{c})$ and $\eta_{\pi}=0.25$. The inset defines the scattering processes: ``(0,0)'' and ``$(\pi, \pi)$'' correspond to a region $\sqrt{q_{x}^{2}+ q_{y}^{2}} \leq 2 \pi \eta_{0}$ and $\sqrt{( |q_{x}|-\pi )^{2} + ( |q_{y}|-\pi )^{2}} \leq 2 \pi \eta_{\pi}$, respectively, and ``others'' denotes the rest of scattering processes. (b) Momentum dependence of the static spin susceptibility at two temperatures. 
}
\label{process}
\end{figure}

We first focus on the retardation effect and discard $V^{*}(\vq)$ in \eq{Vhat}. Since scattering processes are crucially important to the self-restraint effect, we first define them as shown in the inset of \fig{process}(a): ``(0,0)''---scattering processes specified by the momentum transfer $| \vq | \leq 2 \pi \eta_{0}$, ``$(\pi, \pi)$''---those specified by $\sqrt{( |q_{x}|-\pi )^{2} + ( |q_{y}|-\pi )^{2}} \leq 2 \pi \eta_{\pi}$, and ``others''---the other  scattering processes; here $\vq$ is defined in the first Brillouin zone.  We then compute Eqs.~(\ref{eliashberg1}) and (\ref{eliashberg2}) at $T=0.01$ by varying $\eta_{0}$ and $\eta_{\pi}$ for ``(0,0)'', ``$(\pi,\pi)$'', ``others'', and also for including all scattering processes denoted by ``all''. 

In \fig{process}(a), we fix $\eta_{\pi}=0.25$ and study the $\eta_{0}$ dependence of the eigenvalue. No  $\eta_{0}$ dependence occurs for ``all'' and ``$(\pi, \pi)$'' as it should be. Since antiferromagnetic fluctuations  have large spectral weight around $\vq = (\pi, \pi)$, it is natural to have the largest eigenvalue for ``$(\pi, \pi)$''. This clearly indicates that the superconductivity is indeed driven by spin fluctuations around $\vq=(\pi, \pi)$. 

The key result is that the eigenvalue for ``$(\pi, \pi)$'' is suppressed by including ``(0,0)'' (see the results for ``(0,0)''+``$(\pi,\pi)$'') and results for ``all'' is reproduced for $\eta_{0}\gtrsim 0.25$. Given that the spin susceptibility is very small in a small $\vq$ region as shown in \fig{process}(b) and thus naturally the eigenvalue for ``(0,0)'' is very small in \fig{process}(a), it is a giant effect that the eigenvalue for ``$(\pi, \pi)$'' is suppressed by more than 15 \% by the scattering processes of ``(0,0)''. We call this suppression the {\it self-restraint effect} of superconductivity. The self-restraint effect can also be seen in similar calculations by fixing $\eta_{0}=0.25$ and changing $\eta_{\pi}$ as shown in Appendix~A. Note that there is not a peak structure at $\vq=(0,0)$ in \fig{process}(b) and thus the scattering processes of ``(0,0)'' should not be confused with ferromagnetic fluctuations competing with the singlet pairing instability.

\subsubsection{Phase diagram}
\begin{figure}[ht]
\centering
\includegraphics[width=8cm]{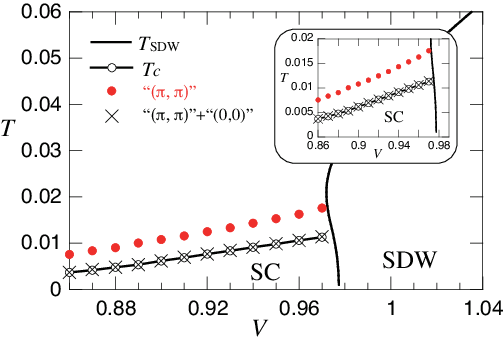}
\caption{Phase diagram of superconductivity and SDW in the plane of the interaction strength $V$ and temperature $T$. ``$(\pi, \pi)$'' and ``$(0, 0)$''+``$(\pi, \pi)$'' indicate $T_{c}$ obtained by considering only scattering processes specified by ``$(\pi, \pi)$'' and those by ``$(0, 0)$'' and ``$(\pi, \pi)$'', respectively.  The inset magnifies the superconducting phase. 
}
\label{phase}
\end{figure}

How much is the actual $T_{c}$ suppressed by the self-restraint effect? Since the effect from the ``(0,0)'' and ``$(\pi, \pi)$'' scattering processes  saturates in $\eta_{0} \geq 0.25$ [\fig{process}(a)] and $\eta_{\pi} \geq 0.25$ (Appendix~A),  we choose $\eta_{0}=0.25$ and $\eta_{\pi}=0.25$ to define each scattering process. The value of $\eta_{\pi}=0.25$ roughly corresponds to the region where the spin susceptibility is large---this region shows a weak temperature dependence as seen in \fig{process}(b), although the peak height depends on temperature. Performing comprehensive computations by changing $T$ and the interaction strength $V$, we construct the phase diagram as shown in \fig{phase}. The SDW phase is realized for a large $V$ as expected and its transition always occurs at $\vq=(\pi, \pi)$ along the whole critical line. The reentrant behavior of the critical line at low $T$ around $V=0.97$--$0.98$ is due to the sharpening of the Fermi surface. The critical temperature of superconductivity $T_{c}$ increases monotonically upon approaching the SDW phase because of the enhancement of spin fluctuations. $T_{c}$ obtained only from the ``$(\pi, \pi)$'' scattering is suppressed nearly by half by including the ``(0,0)'' scattering, with which $T_{c}$ from all scattering processes is reproduced. That it, as already implied in \fig{process}(a), the spin-fluctuation mechanism of superconductivity  suffers from a big self-restraint effect---$T_{c}$ is suppressed nearly by half with a seemingly tiny effect from the ``(0,0)'' scattering process.  

\subsubsection{Pairing gap and renormalization function}
\begin{figure}[ht]
\centering
\includegraphics[width=7cm]{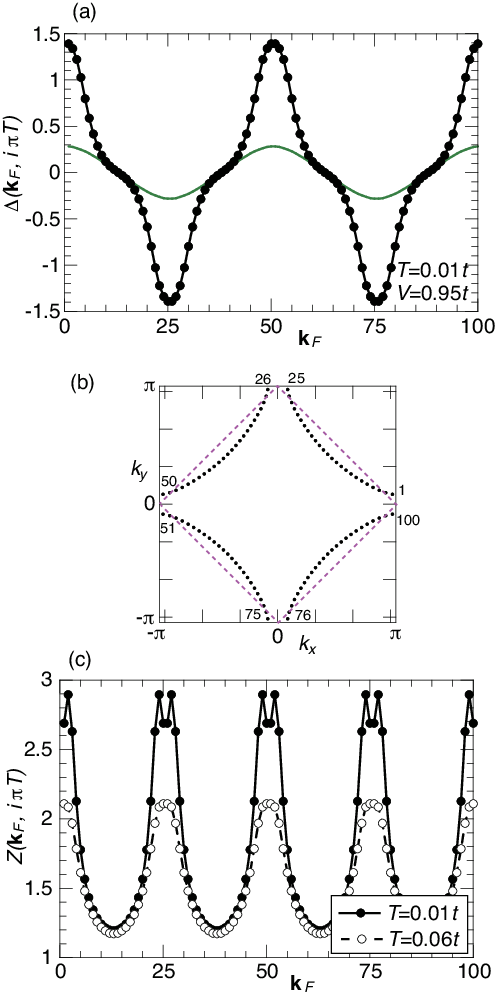}
\caption{Momentum dependence of the pairing gap and  the renormalization function at the lowest Matsubara frequency along the Fermi surface. (a) The pairing gap at $T=0.01(\approx T_{c})$. The curve in green describes $\Delta(\vk_{F}) = - 0.144(\cos k_{x} - \cos k_{y})$. The Fermi momenta $\vk_{F}$ is numbered from 1 to 100 as shown in (b); the dashed line denotes the magnetic zone boundary. The ``$(\pi,\pi)$'' scattering processes are optimized between two $\vk_{F}$ points: 2-74, 24-52, 27-99, and 49-77. (c) The renormalization function  at $T=0.01$ and $0.06$.  
}
\label{gap}
\end{figure}

Figure~\ref{gap}(a) shows the momentum dependence of the pairing gap at the lowest Matsubara frequency. The gap is characterized by the $B_{1}$ representation of the $C_{4v}$ point group symmetry, the so-called $d_{x^{2}-y^{2}}$-symmetry. The key insight here is that the gap is not characterized  by the simple form such as $\Delta(\vk) \propto \cos k_{x} - \cos k_{y}$, as shown in the green curve fitted to the nodal regions. This means that the pairing gap contains higher harmonics with $d$-wave symmetry. Consequently, the gap tends to be enhanced substantially around $\vk \approx (\pi, 0)$ and $(0, \pi)$. This is reasonable because spin fluctuations are maximized when two momenta on the Fermi surface are connected with $\vq =(\pi, \pi)$ as seen in \fig{gap}(b). 

Figure~\ref{gap}(c) is the momentum dependence of the renormalization function $Z$ at the lowest Matsubara frequency. As expected, $Z$ becomes large around $\vk =(\pi, 0)$ and  $(0, \pi)$. In particular, when the system is very close to the SDW phase, low-energy spin fluctuations have a very sharp peak at $\vq=(\pi, \pi)$, which then leads to a dip structure in $Z$ at the zone boundary where the scattering wavevector of spin fluctuations deviates slightly from $\vq=(\pi, \pi)$ [see also \fig{gap}(b)]. This dip structure vanishes upon going away from the SDW phase---for example by increasing $T$---and the resulting $Z$ exhibits a single peak at the zone boundary; see the result at $T=0.06t$ in \fig{gap}(c). 

We checked that the self-restraint effect does not yield a sizable change in the momentum dependence of $\Delta$ and $Z$ by performing similar calculations with the ``$(\pi,\pi)$'' scattering alone (see Appendix~B).  This is different from results obtained in the two-band model, where the so-called $s\pm$-wave symmetry is realized and the self-restraint effect leads to a large momentum dependence of the pairing gap \cite{yamase20}.  

\subsubsection{Intuitive understanding of the self-restraint effect} 
The self-restraint effect is understood intuitively by considering phase frustration of the pairing gap. As is well known, spin fluctuations yield a repulsive pairing interaction---$\Gamma_{\vk_{F} \vk_{F}'}$ is positive in \eq{eliashberg1}. Hence to fulfill \eq{eliashberg1} the pairing gap has the opposite sign between two $\vk_{F}$ points connected by spin fluctuations. This is the very reason why the $d$-wave pairing gap, more broadly a gap with the sign change, is preferred in the spin-fluctuation mechanism. However, there always remains some weight of spin fluctuations even far away from $\vq=(\pi, \pi)$ as a {\it tail} of spin fluctuations. In particular, spin fluctuations with small $\vq$ tend to change the sign of the pairing gap when two momenta are close to each other on the Fermi surface. This necessarily leads to frustration  of the pairing gap since the $d$-wave gap has the same sign when two momenta are nearby except for the nodal region. A common belief is that such frustration could be negligible because spin fluctuations with small $\vq$ are very weak. This could be  a major reason why the self-restraint effect has not been recognized in the long history of the Eliashberg theory. However, such a belief is not correct. Rather the phase frustration of the pairing gap can become sizable to suppress the tendency of superconducting instability as shown in \fig{phase} even though fluctuations with small $\vq$ are seemingly very weak---note that the eigenvalues themselves become very small only for ``(0,0)'' as expected (see \fig{process}). 

On the other hand, the self-restraint effect cannot occur for an attractive pairing interaction, for example, electron-phonon coupling \cite{bardeen57,schrieffer} and orbital nematic fluctuations \cite{yamase13b}, because the pairing gap tends to have the same sign, which does not lead to any frustration---all scattering processes work constructively to the superconducting instability. Hence we conclude that the self-restraint effect is an intrinsic feature of the repulsive pairing interaction, which necessarily leads to phase frustration of the pairing gap. This explanation does not depend on details of models and approximations.

\subsection{Role of instantaneous magnetic interaction} 
Next we consider the effect of $V^{*}(\vq)$ in \eq{Vhat}, which does not have a frequency dependence and accounts for an instantaneous magnetic interaction. A subtle aspect here is that the Coulomb interaction is also instantaneous and in general is expected to suppress  the effect of $V(\vq)$. We thus model the instantaneous interaction as 
\be
V^{*}(\vq) = V_{0} + r  V(\vq) \,.
\label{V*} 
\ee
Here $V_{0}$ is the effect from the on-site Coulomb interaction, which may play a role of the pseudo-Coulomb interaction $\mu^{*}$ \cite{schrieffer} in the energy scale of the magnetic interaction. However, as far as the resulting pairing symmetry is of $d$-wave, which is the case of the present work, the effect of $V_{0}$ is cancelled after the momentum summation in \eq{eliashberg1}. A new aspect is that we then consider the nearest-neighbor Coulomb interaction, which can be modeled by the parameter $r (<1)$---it is positive and its smaller value means a larger suppression of the pairing interaction. 

Solving Eqs.~(\ref{eliashberg1}) and (\ref{eliashberg2}) by including the instantaneous interaction  $V^{*}(\vq)$, we need to introduce a cutoff energy of Matsubara frequency in \eq{eliashberg1} as is well known in literature \cite{morel62,schrieffer,marsiglio20}; the instantaneous term does not contribute to the right hand side of \eq{eliashberg2}. We choose the cutoff $i\omega_{c} \approx i 20 t$, whose magnitude is two times larger than the band width. Since this choice of $\omega_{c}$ is not unique, $r$ should be regarded as a phenomenological parameter to describe the suppression of the pairing instability by the nearest-neighbor Coulomb repulsion. This technical aspect, however, does not affect our conclusions. 

\begin{figure}[ht]
\centering
\includegraphics[width=7cm]{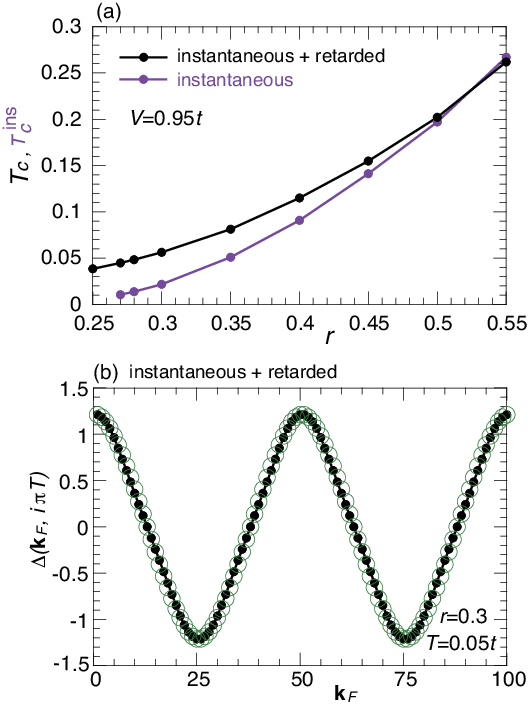}
\caption{Results in the presence of the instantaneous interaction. (a) $T_{c}$ as a function of $r$---the effect of the Coulomb repulsion [see \eq{V*}] in two different cases: only the instantaneous interaction, i.e., only $V^{*}(\vq)$ in \eq{Vhat}, and both instantaneous and retarded interactions, i.e., the full expression of \eq{Vhat}. (b) Momentum dependence of the pairing gap at the lowest Matsubara frequency at $T=0.05 (\approx T_{c})$ along the Fermi surface [see \fig{gap}(b)] when both instantaneous and retarded interactions are considered. The green circles correspond to the simple $d$-wave gap described by $\Delta(\vk) = -1.237\,(\cos k_{x} - \cos k_{y})/2$. 
}
\label{phasei3}
\end{figure}

Figure~\ref{phasei3}(a) is $T_{c}$ as a function of $r$. As expected, $T_{c}$ is suppressed substantially by decreasing $r$, namely increasing the Coulomb repulsion, and reproduces $T_{c}=0.0098t$ at $V=0.95t$ in the limit of $r=0$,  where only the retardation effect remains. A key point here is that the instantaneous interaction $V^{*}(\vq)$ can boost $T_{c}$ substantially and the actual value of $T_{c}$ depends strongly on the effect of the Coulomb repulsion described by $r$. 

In contrast to \fig{gap}(a), the pairing gap is characterized by the simple form of $\cos k_{x} - \cos k_{y}$ even for a reasonably small instantaneous interaction, for example at $r=0.3$, as shown in \fig{phasei3}(b). This means that the pairing gap is formed dominantly by the nearest-neighbor interaction in real space---a direct proof that the momentum dependence of $\Delta(\vk_{F})$ comes from the instantaneous interaction $V(\vq)$, which occurs for the nearest-neighbor sites in real space. 

The renormalization function, on the other hand, is essentially the same as \fig{gap}(c) where only the retarded interaction is considered. This is because the instantaneous part is cancelled after the Matsubara summation in \eq{eliashberg2}, indicating that the momentum dependence of $Z$ is determined by the retarded interaction. See Appendix~C for actual results of the momentum dependence of $Z$. 

While the effect of the instantaneous interaction is dominant even for a small $r$, this does not mean that the retardation effect is irrelevant. In \fig{phasei3}(a), we also plot  $T_{c}^{\rm ins}$ obtained by considering only the instantaneous interaction. $T_{c}^{\rm ins}$  becomes less than 0.01 in $r \lesssim 0.3$, but $T_{c}$ retains a value a few times higher than $T_{c}^{\rm ins}$. This implies that both retardation and instantaneous interactions work constructively to achieve a relatively high $T_{c}$ especially when the instantaneous interaction is suppressed by the Coulomb repulsion, i.e., in the small $r$ region in \fig{phasei3}(a). 

It should be noted that these results rely on the special form of $V(\vq)$, not a general feature of the instantaneous interaction. $V(\vq)$ works as a repulsive pairing interaction for $\vq \approx (\pi, \pi)$ and an attractive one for a small $\vq$ because of the sign change of $V(\vq)$---note that the retarded interaction is always repulsive [see \eq{Vhat}].  Hence this specific instantaneous interaction does not yield phase frustration of the pairing gap, and is free from the self-restraint effect---all scattering processes work constructively in contrast to \fig{process}. 

If we consider a different from of $V(\vq)$, for example, the Lorentz-type interaction which is repulsive for any $\vq$, the instantaneous interaction suffers from the self-restraint effect. We find that this effect is huge and the substantial enhancement of $T_{c}$ does not occur (see Appendix~D).  

\section{Conclusions and discussions}
In the framework of the Eliashberg theory, we have analyzed the spin-fluctuation mechanism of superconductivity on a square lattice. The key insight is the importance of the self-restraint effect of superconductivity, which suppresses $T_{c}$ substantially. The self-restraint effect can be understood in terms of phase frustration of the pairing gap induced by a repulsive pairing interaction, typical to the spin-fluctuation mechanism. The self-restraint effect, however, can be circumvented by a special form of the instantaneous interaction. A plausible interaction is $V(\vq) \propto ( \cos q_{x} + \cos q_{y})$, which can lead to $T_{c}$ one order magnitude higher than the retardation effect even if the effect of the Coulomb interaction is considered appropriately. For example, we may  achieve high $T_{c} (= 0.06t)$ even at $r=0.3$ in \fig{phasei3}(a), whereas without the instantaneous interaction, the highest $T_{c}$ is around $T_{c} \sim 0.01t$ (\fig{phase}), which may corresponds to around 20 K at most if $t$ is around 100-200 meV.  While vertex corrections are neglected in the Eliashberg theory, explicit studies found that the vertex corrections reduce the pairing instability \cite{hotta94,huang06,kitatani19}. In this sense, we may interpret our obtained $T_{c}$ as the upper limit. A low value of $T_{c} (\sim 0.01 t)$ due to the retardation effect is broadly consistent with many studies \cite{shimahara88,bickers89,radtke92,hotta94,lenck94,pao94,huang06,kitatani19}. Furthermore Ref.~\onlinecite{dong22} reported that spin-fluctuation theory alone cannot account for the superconductivity in the Hubbard model. 

While the spin-fluctuation mechanism of superconductivity has been widely studied in literature \cite{scalapino12}, neither the presence of the self-restraint effect nor the important role of the instantaneous interaction has been recognized. To see the former effect, one needs to compute the eigenvalue by defining the scattering processes. This kind of analysis was not performed usually, except for Ref.~\onlinecite{dong22,dong22a}, where small momentum fluctuations were observed to yield a negative contribution to the anomalous self-energy. To see the latter effect, we need to separate contributions from the instantaneous and retarded interactions. While Ref.~\onlinecite{maier08} noticed such a separation, the conclusions that we have reached were not reported. 

The pairing gap tends to be largely deviated from the simple $d$-wave form factor when we focus on the retarded interaction [\fig{gap}(a)]. This feature is not a special feature of the Eliashberg theory \cite{lenck94}, but similar results were also obtained in functional renormalization group studies \cite{reiss07,jwang14} as well as second-order renormalized perturbation theory \cite{neumayr03}. Intuitively, this is easily understood. The simple $d$-wave pairing gap implies that a pairing interaction is dominant between the nearest-neighbor sites in real space. Hence \fig{phasei3}(b) is considered as the typical result due to the dominant pairing interaction such as $V(\vq) \propto \cos k_{x} + \cos k_{y}$. On the other hand,  the spin fluctuations usually lead to weaker, but non-negligible pairing interactions beyond the nearest-neighbor sites in real space, yielding higher harmonics in the $d$-wave symmetry in momentum space as shown in \fig{gap}(a). 

In experiments, the momentum dependence of the $d$-wave superconducting gap in cuprates follows almost perfectly the function of $\cos k_{x} - \cos k_{y}$, not only in the optimally-doped and overdoped regions  \cite{ding96,sdchen22} but also in the underdoped region if the pseudogap effect is considered separately \cite{yoshida12}. This implies that the instantaneous interaction $V(\vq) \propto  ( \cos q_{x} + \cos q_{y})$---free from the self-restraint effect---plays the crucial role to the mechanism of high $T_{c}$ in cuprates. 

Although the issue of the pseudogap is beyond the scope of the present work, our \fig{gap}(a) shows that the retarded spin fluctuations generate a large gap especially around $(\pi, 0)$ and $(0, \pi)$. This result might be useful when one considers the origin of the pseudogap in cuprates, which is also enhanced around the same momenta. 

Anderson highlighted the vital role of the instantaneous pairing interaction rather than the retarded spin fluctuations to understand the high-$T_{c}$ cuprate superconductors \cite{anderson07}. This insight is in line with the present work as long as the Coulomb repulsion suppresses moderately the effect of the instantaneous pairing interaction. When the Coulomb repulsion suppresses it severely, for example, in $r \lesssim 0.3$ in \fig{phasei3}(a), both retardation and instantaneous interactions are necessary to attain high $T_{c}$. We should note, however, that the present theory is based on the Eliashberg theory and thus strong correlation effect is not considered properly except that the strong on-site Coulomb repulsion is cancelled out. Application of the present work to cuprates may be limited to the overdoped region, where correlation effects are believed to be weakened. 

As shown in \fig{phasei3}(a), $T_{c}$ strongly depends on the degree of the Coulomb repulsion. The effect of the Coulomb interaction may depend on materials, which can be an important factor to the material dependence of $T_{c}$ in the cuprate-superconductor family. Since the material dependence of $T_{c}$ is frequently discussed in terms of the magnitude of $t'$ \cite{pavarini01} and the number of CuO$_{2}$ layers \cite{iyo07}, it is very interesting to study how those are related to the effect of the Coulomb repulsion.

From a view of the present theory, a special feature of cuprate superconductors lies in the special form of $V(\vq)$, which makes it possible to achieve high $T_{c}$. The spin-fluctuation mechanism of superconductivity is believed to  be general and applicable also to other systems \cite{scalapino12}, which are characterized by a much lower $T_{c}$ than that of cuprates. This difference may come from a different form of $V(\vq)$. For example, as shown in Appendix~D, the Lorentz-type and Hubbard-like interactions may be invoked for those systems, which can reproduce a relatively low $T_{c}$. 

We have discussed superconductivity from a view of magnetic interactions. The obtained insight however can be operative as long as a pairing interaction becomes repulsive in a general situation because it necessarily yields phase frustration of pairing gap---the self-restraint effect. On the other hand, the self-restraint effect does not occur if the pairing interaction becomes attractive. This may occur for, for example, electron-phonon interactions  \cite{bardeen57,schrieffer,eliashberg60,marsiglio20,ashcroft68,ashcroft04} and charge interactions \cite{perali96} including orbital fluctuations \cite{stanescu08,kontani11} and electronic nematic fluctuations \cite{yamase13b,agatsuma16}. In this case, all scattering processes can contribute to superconductivity constructively and provide a favorable situation to achieve possibly high $T_{c}$.

\acknowledgments
The author thanks W. Metzner for a critical reading of the manuscript and valuable comments,  
and A. Greco for enlightening comments including ones about Refs.~\onlinecite{mansart13,omahony22,dong22}. 
He was supported by JSPS KAKENHI Grants No.~JP20H01856.

\appendix

\setcounter{equation}{0}
\renewcommand{\theequation}{A\arabic{equation}}%

\section{Supplement to \fig{process}(a)} 
In \fig{process}(a), we have presented the $\eta_{0}$ dependence of the eigenvalue for each scattering process to demonstrate the self-restraint effect. Here we present the $\eta_{\pi}$ dependence of the eigenvalue by fixing $\eta_{0}=0.25$. As expected, the eigenvalues for ``(0,0)'' and ``all'' do not depend on $\eta_{\pi}$, whereas the eigenvalue for ``$(\pi, \pi)$'' increases with increasing $\eta_{\pi}$. The self-restraint effect is recognized as the suppression of the eigenvalue from ``$(\pi, \pi)$'' by including the ``(0,0)'' scattering processes---results for ``(0,0)''+``$(\pi, \pi)$''---and the eigenvalue for all scattering processes is reproduced in $\eta_{0} \geq 0.25$.

\begin{figure}[ht]
\centering
\includegraphics[width=7cm]{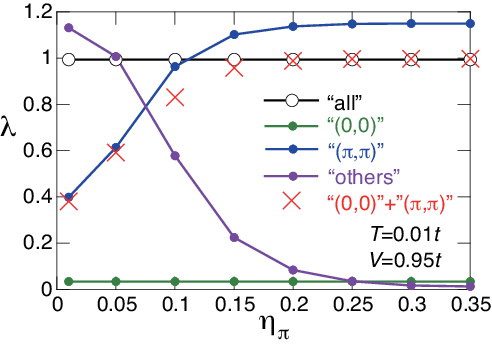}
\caption{$\eta_{\pi}$ dependence of the eigenvalue $\lambda$ for different scattering processes at $T=0.01(\approx T_{c})$ and $\eta_{0}=0.25$. }
\label{process-Supp}
\end{figure}

\section{Self-restraint effect on $\boldsymbol{\Delta}$ and $\boldsymbol{Z}$} 
\begin{figure}[ht]
\centering
\includegraphics[width=7cm]{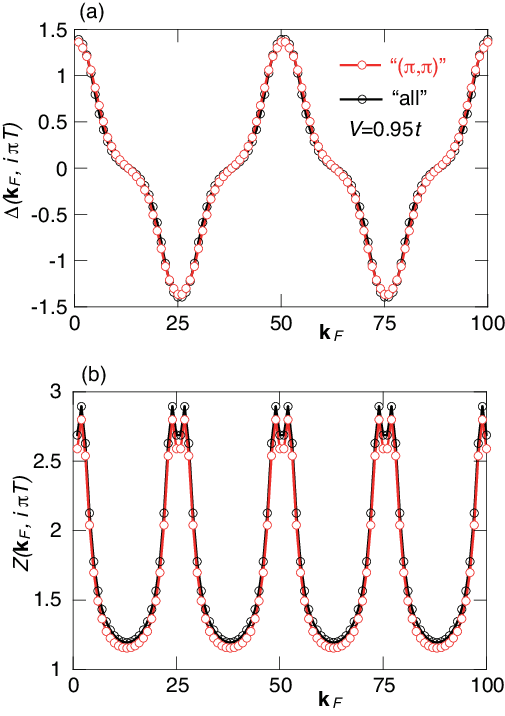}
\caption{Momentum dependence of the pairing gap (a) and the renormalization function (b) at the lowest Matsubara frequency along the Fermi surface at $T=0.01 (\approx T_{c})$ by considering only the scattering processes ``$(\pi, \pi)$''; see \fig{gap}(b) for the definition of $\vk_{F}$. The results are compared with those obtained for the full calculations (``all''). 
}
\label{gap-Supp}
\end{figure}

In Figs.~\ref{gap}(a) and (c) we have presented the pairing gap and the renormalization function, respectively. In \fig{gap-Supp}, we superimpose results from the ``$(\pi, \pi)$'' scattering alone. There is no important difference between ``$(\pi, \pi)$'' and ``all'', indicating that the self-restraint effect does not affect the momentum dependence of $\Delta(\vk_{F})$ and $Z(\vk_{F})$.

\section{Effect of instantaneous interaction on $\boldsymbol{Z}$} 
\begin{figure}[ht]
\centering
\includegraphics[width=7cm]{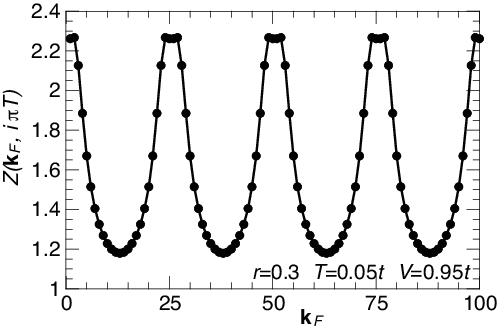}
\caption{Momentum dependence of the renormalization function at the lowest Matsubara frequency along the Fermi surface [see \fig{gap}(b)] at $T=0.05 (\approx T_{c})$ by considering both instantaneous and retarded interactions; $r=0.3$ is chosen in \eq{V*}. }
\label{phasei3-Supp}
\end{figure}

In \fig{phasei3}(b), we have presented the momentum dependence of $\Delta(\vk_{F})$ in the presence of the instantaneous pairing interaction. For completeness, we also present $Z(\vk_{F})$ in \fig{phasei3-Supp} for the same parameters as those in \fig{phasei3}(b). Qualitatively, it shows essentially the same feature as \fig{gap}(c), where only the retardation effect is considered. However, temperature here is $T=0.05$, higher than $T=0.01$ in \fig{gap}(c), and thus thermal broadening is large enough to smear a dip near the zone boundary. 

\section{Different choices of $\boldsymbol{V(\vq)}$}  
In the main text we have considered the magnetic interaction $V(\vq)=2V(\cos q_{x} + \cos q_{y})$. In real space, this means that the dominant magnetic interaction occurs between the nearest-neighbor sites. Here, we may consider two other choices: i) a Lorentz-type interaction and ii) a Hubbard-like interaction. 

\subsection{Lorentz-type interaction} 
A Lorentz-type interaction describes an exponential-like decay of the interaction in real space and may be given in momentum space by 
\be
V_{L} (\vq) = -2 V_{L} \sum_{n, m} \frac{\Gamma}{(\vq - \vQ_{n m})^{2} + \Gamma^2} \, , 
\label{Lorentz} 
\ee
where $V_{L}$ controls the magnitude of interaction, $\Gamma$ describes the peak width, and $\vQ_{n m}=(\pi+ 2n\pi, \pi+2m\pi)$ with $n$ and $m$ being integers. We have added the subscript ``$L$'' to $V_{L}(\vq)$, not to mix up with the results for $V(\vq)$ given in the main text. To achieve the symmetric distribution around $\vq=(\pm\pi, \pm\pi)$, we consider $n=0, \pm 1, \pm 2, \pm 3, \pm4, -5$ and  $m=0, \pm 1, \pm 2, \pm 3, \pm4, -5$. We fix $\Gamma =1$. 

\begin{figure}[ht]
\centering
\includegraphics[width=7cm]{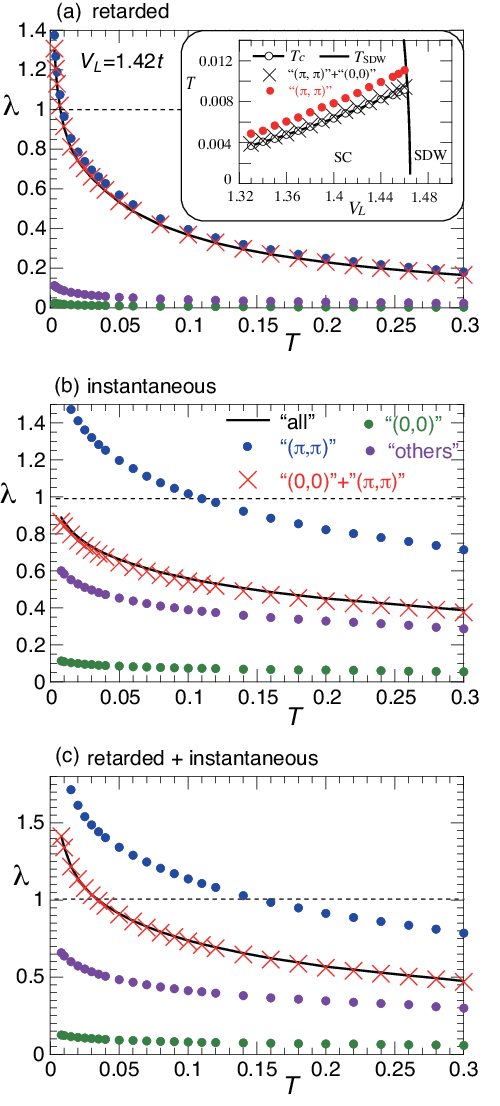}
\caption{Temperature dependence of the eigenvalue $\lambda$ for each scattering process. (a) Only the retarded interaction is considered. The inset shows the phase diagram, where $T_{c}$ obtained from ``$(\pi, \pi)$'' only is also shown to clarify the self-restraint effect. (b) Only the instantaneous interaction is considered. (c) Both retarded and instantaneous interactions are considered. 
}
\label{L-lambda}
\end{figure}

After checking that we obtained results similar to Figs.~\ref{process} and \ref{process-Supp}, we choose $\eta_{0}=\eta_{\pi}=0.25$, the same parameters as the main text. Figure~\ref{L-lambda}(a) is the temperature dependence of the eigenvalue $\lambda$ only for the retarded interaction for each scattering process at $V_{L}=1.42$. The eigenvalue is the largest for ``$(\pi, \pi)$'', increases with decreasing temperature, and exceeds unity at low temperature, indicating that the superconductivity is driven by the antiferromagnetic fluctuations with the momentum transfer $(\pi,\pi)$. Naturally, the eigenvalue for ``$(0,0)$'' is the smallest and its contribution seems negligible. However, when we include both ``$(\pi, \pi)$'' and ``(0,0)'' scattering processes, the eigenvalue is suppressed---the self-restraint effect---and reproduce results obtained for all scattering processes. The inset in \fig{L-lambda}(a) shows the phase diagram, where we also plot $T_{c}$ obtained only for the ``$(\pi, \pi)$'' scattering to clarify how much $T_{c}$ is suppressed by the self-restraint effect. $T_{c}$ is suppressed approximately 20 \% by the ``(0,0)'' scattering. While the phase diagram itself is essentially the same as \fig{phase}, including the SDW phase, the self-restraint effect becomes smaller for the magnetic interaction given by \eq{Lorentz} than $V(\vq)$ employed in the main text. The reason for this is easily pinned down by looking at the absolute value of $\lambda$ for the ``(0,0)'' scattering at $T_{c}$: $\lambda=0.019$ and $0.035$ for $V_{L}(\vq)$ [\fig{L-lambda}(a)] and $V(\vq)$ [\fig{process}(a)], respectively.  This means that the effect of ``$(0,0)$'' scattering on the superconductivity is smaller for $V_{L}(\vq)$ and so is the self-restraint effect.

Next we focus on the instantaneous interaction, i.e., the first term on the right hand side in \eq{Vhat}, where $V^{*}(\vq)=V_{L}(\vq)$. Although both $V_{L}(\vq)$ and $V(\vq)$ have a peak at $\vq=(\pi, \pi)$ with a minus sign favoring a SDW state with $(\pi, \pi)$,  $V_{L}(\vq)$ is qualitatively different from $V(\vq)$ in that $V_{L}(\vq)$ is negative definite whereas $V(\vq)$ changes its sign for small $\vq$. The obtained temperature dependence of $\lambda$ is shown in \fig{L-lambda}(b). $\lambda$ for the ``$(\pi, \pi)$'' scattering crosses unity around $T=0.106$---a favorable situation of high $T_{c}$. However, once we include the ``$(0,0)$'' scattering (see results for ``(0,0)''+``$(\pi, \pi)$''), $\lambda$ is suppressed substantially and does not cross unity. The resulting $\lambda$ almost reproduces results for ``all''. Compared with the case of the retarded interaction [\fig{L-lambda}(a)] the self-restraint effect is remarkably large. This is because $\lambda$ for the ``(0,0)'' scattering is relatively large to be $\lambda=0.11$ at the lowest temperature, indicating that the phase frustration effect caused by the ``(0,0)'' scattering processes is pronounced more than \fig{L-lambda}(a). 

In \fig{L-lambda}(c) we compute $\lambda$ by including both retarded and instantaneous interactions. They work constructively and lead to the superconducting instability at $T_{c}=0.035$, where $\lambda$ exceeds unity. Nonetheless, given that $\lambda$ exceeds unity at $T=0.15$ for the ``$(\pi, \pi)$'' scattering alone, the self-restraint effect is huge. This is because $\lambda$ only for the ``$(0,0)$'' scattering is relatively large, i.e., $\lambda=0.099$  at $T_{c}$, and thus the superconducting instability is severely suppressed by phase frustration caused by the ``(0,0)'' scattering. 

\begin{figure}[ht]
\centering
\includegraphics[width=7cm]{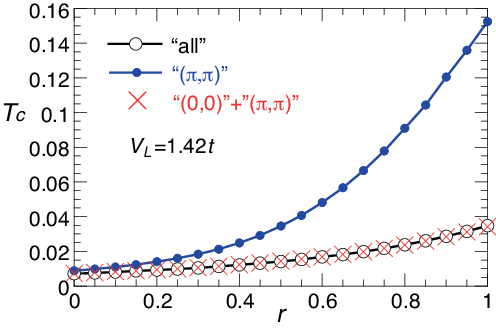}
\caption{Suppression of $T_{c}$ by the Coulomb repulsion. The effect of Coulomb repulsion is controlled by $r$ in \eq{V*}; the effect of $V_{0}$ is cancelled for the present $d$-wave pairing gap. Results for the ``$(\pi, \pi)$'' and ``$(0,0)$''+``$(\pi, \pi)$'' scattering processes are also shown to highlight the importance of the self-restraint effect. 
}
\label{L-Vphase}
\end{figure}

\fig{L-lambda} suggests that while both retarded and instantaneous interactions suffer from the self-restraint effect,  the effect of the instantaneous interaction is vital to achieving high $T_{c}$. The instantaneous interaction, however, may be renormalized by the Coulomb repulsion as we have discussed in the main text. Employing the renormalized interaction \eq{V*}, we study how the superconducting instability is suppressed. Note that the effect of $V_{0}$ in \eq{V*} is cancelled out as long as the pairing gap is of $d$-wave, which is the case in the present study. The $r$ dependence of $T_{c}$ is presented in \fig{L-Vphase}. $T_{c}$ is suppressed monotonically with reducing $r$, namely with increasing the effect of the Coulomb repulsion. Notably the $r$ dependence is much weaker than \fig{phasei3}(a). 

In \fig{L-Vphase}, we also plot results of $T_{c}$ only from the ``$(\pi, \pi)$'' scattering and those from the ``$(0,0)$"+``$(\pi,\pi)$" scattering. The huge suppression of $T_c$ by including the ``$(0,0)$" scattering manifests the importance of the self-restraint effect due to the instantaneous interaction especially for a large $r$. While the self-restraint effect might seem to disappear at $r=0$---the instantaneous part vanishes---in the scale of \fig{L-Vphase}, the self-restraint effect from the retardation part still gives the suppression of $T_{c}$ approximately 20 \% as seen in \fig{L-lambda}(a).

\begin{figure}[ht]
\centering
\includegraphics[width=14cm]{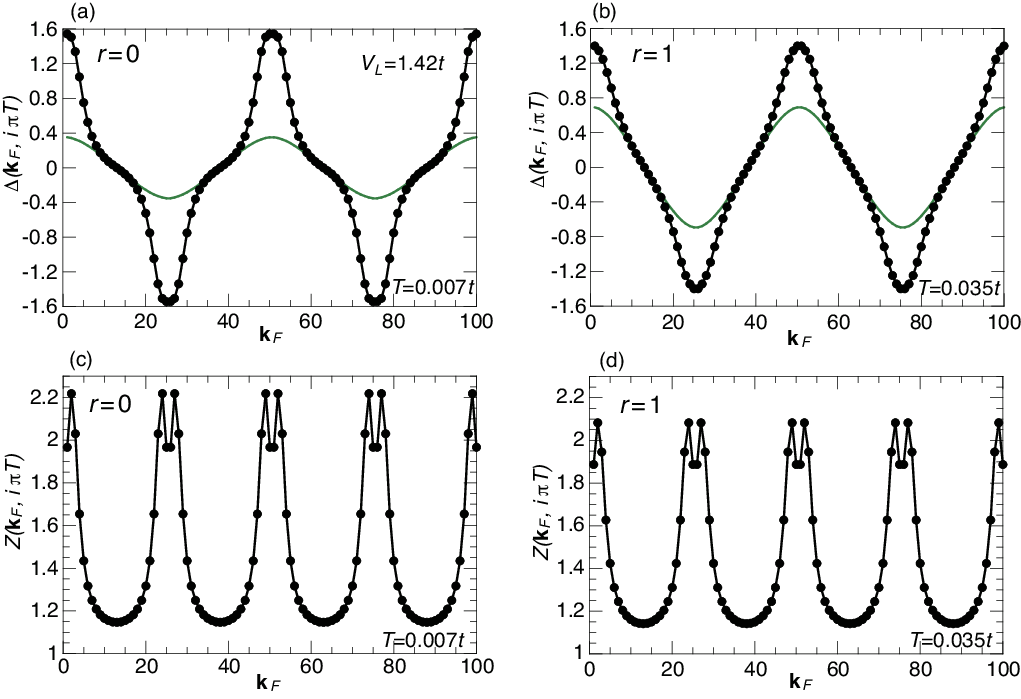}
\caption{Momentum dependence of the pairing gap $\Delta$ and the renormalization function $Z$ along the Fermi surface at the lowest Matsubara frequency; see \fig{gap}(b) for the definition of $\vk_{F}$. (a) and (c) Only the retarded interaction ($r=0)$ is considered at $T=0.007 (\approx T_{c})$. (b) and (d) Both retarded and instantaneous interactions $(r=1)$ are considered at $T=0.035 (\approx T_{c})$.  The simple $d$-wave gap described by $\Delta(\vk) = -\Delta_{0}\,(\cos k_{x} - \cos k_{y})/2$ is also shown in the green curve in (a) and (b) with  $\Delta_{0}=0.3574$ and $0.7057$, respectively. 
}
\label{L-gap}
\end{figure}

We also study the momentum dependence of the pairing gap and the renormalization function. Figures~\ref{L-gap}(a) and (b) show $\Delta(\vk_{F})$ for $r=0$ and $1$, respectively: $r=0$ corresponds to the case of the retarded interaction only and $r=1$ the case of both retarded and instantaneous interactions. $\Delta(\vk_{F}
)$ exhibits the so-called $d$-wave symmetry, but deviates from the simple form such as $\cos k_{x} - \cos k_{y}$. This is because the pairing interaction extends beyond the nearest-neighbor sites in real space, yielding higher harmonics. A comparison between Figs.~\ref{L-gap}(a) and (b) shows that the enhancement of the gap around $(\pi,0)$ is pronounced more for the retarded interaction. Consequently $\Delta(\vk_{F})$ deviates the simple $d$-wave form more largely. On the other hand, the momentum dependence of $Z(\vk_{F})$ is essentially the same for both $r=0$ and 1 as shown in Figs.~\ref{L-gap}(c) and (d), respectively. $Z(\vk_{F})$ becomes the smallest around the nodal region $\vk \approx (0.4\pi, 0.4\pi)$ and is enhanced around $\vk \approx (\pi, 0)$. The dip occurs at the zone boundary, similar to \fig{gap}(c) in the main text.

\subsection{Hubbard-like interaction} 
We have considered the magnetic interaction $V(\vq)$ in the main text and $V_{L}(\vq)$ in \eq{Lorentz}, which surely favor a SDW state characterized by the momentum $\vq = (\pi, \pi)$.  However, one may wish to consider a simpler form such as 
\be
V_{H}(\vq)= - 2 V_{H} 
\label{hubbard}
\ee
because this term is expected to be the leading term of our magnetic interaction in \eq{model} at least from a mathematical point of view. While this interaction does not necessarily favor a SDW state characterized by $\vq = (\pi, \pi)$, we can check that it is stabilized for the present band structure. Hence here we shall present outcome also for the interaction \eq{hubbard}. 

Since $V_{H}(\vq)$ does not depend on momentum, the effect of the instantaneous interaction is cancelled after the momentum summation in \eq{eliashberg1} as long as the $d$-wave pairing is stabilized; see also the discussion below \eq{V*} in the main text. Hence the superconductivity is driven only by the retarded interaction part in \eq{Vhat}. After checking that the SDW phase is stabilized in $V_{H} \gtrsim 1.95$, we  choose $V_{H} =1.9$ and compute the temperature dependence of the eigenvalue in the Eliashberg equations for each scattering process. 

\begin{figure}[ht]
\centering
\includegraphics[width=7cm]{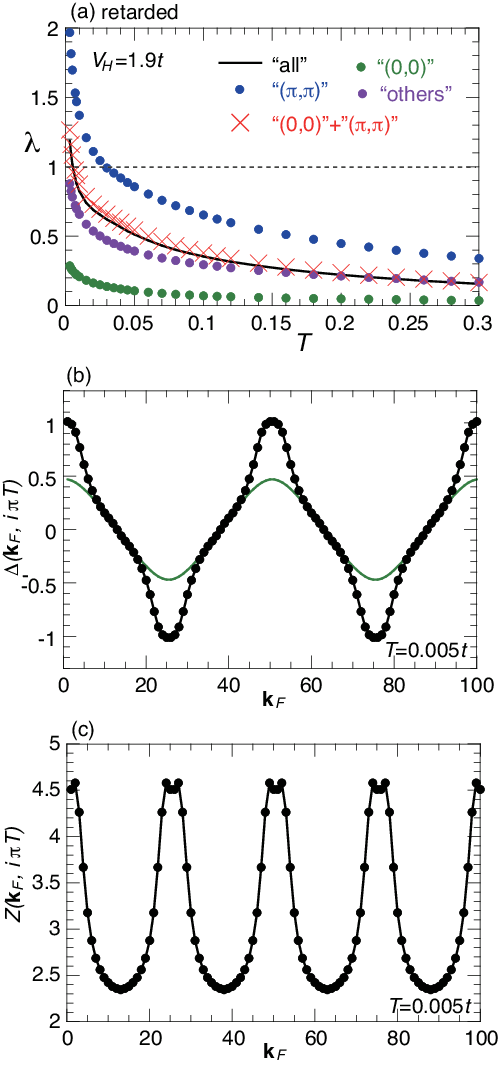}
\caption{Results for the Hubbard-like interaction. (a) Temperature dependence of the eigenvalue $\lambda$ for each scattering process. Only the retarded interaction is considered. (b) Momentum dependence of the pairing gap $\Delta$ along the Fermi surface at the lowest Matsubara frequency; see \fig{gap}(b) for the definition of $\vk_{F}$. The simple $d$-wave gap described by $\Delta(\vk) = -\Delta_{0}\,(\cos k_{x} - \cos k_{y})/2$ is also shown in the green curve with  $\Delta_{0}=0.4787$. (c) Corresponding momentum dependence of the renormalization function $Z$.
}   
\label{H-results}
\end{figure}

Figure~\ref{H-results}(a) shows that the eigenvalue for the ``$(\pi, \pi)$'' scattering processes is the largest and exceeds unity at $T=0.029$, indicating possible superconducting instability from antiferromagnetic fluctuations with the momentum transfer $(\pi,\pi)$.  On the other hand, the effect from the ``(0,0)'' scattering processes is seemingly irrelevant to superconductivity. However, when both ``$(\pi, \pi)$''  and ``(0,0)'' scattering processes are taken into account, the eigenvalue is strongly suppressed---self-restraint effect. The resulting eigenvalue becomes almost the same as that computed by all scattering processes and the superconducting instability eventually occurs at $T_{c}=0.0053$. This means that the self-restraint effect reduces $T_{c}$ more than 80~\%. This huge self-restraint effect comes from a relatively large eigenvalue from the ``(0,0)'' scattering processes, compared to that seen in Figs.~\ref{process}(a) and \ref{L-lambda}(a).

The momentum dependence of the pairing gap is shown in \fig{H-results}(b), which is very similar to that observed in Figs.~\ref{gap}(a) and \ref{L-gap}(a), that it, the paring gap contains higher harmonics and cannot be characterized by a simple form such as $\cos k_{x} - \cos k_{y}$. The momentum dependence of the renormalization function $Z$ is shown in \fig{H-results}(c). The result is very similar to Figs.~\ref{gap}(c) and \ref{L-gap}(c), except that the magnitude of $Z$ is enhanced at all Fermi momenta for the Hubbard-like interaction \eq{hubbard}. This enhancement is interpreted as coming from the enhancement of $\Gamma^{Z}_{\vk_{F} \vk_{F}'} ({\rm i}k_{n}, {\rm i}k_{n}')$ in \eq{eliashberg2} due to the momentum independent term, $V(\vq)^{2}=4 V_{H}^{2}$, in the numerator in the second term in \eq{Vhat}. 

\newpage 

\bibliography{main} 

\end{document}